\documentclass[runningheads]{llncs}
\usepackage{graphicx}
\usepackage{rotating}
\usepackage{xcolor}
\usepackage{url}
\usepackage{footnote}
\usepackage[square,numbers]{natbib}
\usepackage{adjustbox}

\newcommand\blfootnote[1]{%
  \begingroup
  \renewcommand\thefootnote{}\footnote{#1}%
  \addtocounter{footnote}{-1}%
  \endgroup
}

\begin{document}
%
\title{Cross-language Information Retrieval}
\author{Petra Galu\v{s}\v{c}\'{a}kov\'{a}\inst{1}\inst{*}
Douglas W. Oard\inst{2},
Suraj Nair\inst{2}}
\authorrunning{P. Galu\v{s}\v{c}\'{a}kov\'{a} et al.}
%
\institute{$^1$Université Grenoble Alpes, Grenoble, France\\ $^2$University of Maryland, College Park, United States\\
\email{galuscakova@gmail.com}}
\maketitle              
%
%
%

\section{Introduction}
Two key assumptions shape the usual view of ranked retrieval: (1) that the searcher can choose words for their query that might appear in the documents that they wish to see, and (2) that ranking retrieved documents will suffice because the searcher will be able to recognize those which they wished to find.  When the documents to be searched are in a language not known by the searcher, neither assumption is true.  In such cases, Cross-Language Information Retrieval (CLIR) is needed.  This chapter reviews the state of the art for CLIR and outlines some open research questions.

\blfootnote{* The author was affiliated with the University of Maryland at the time of writing.}

The goal of Cross-Language Information Retrieval\footnote{The astute reader might ask why the field is not referred to as ``cross-lingual" information retrieval.  Participants in the SIGIR 1996 workshop voted on a name for the field, and cross-language won.  Language is, however, created through use, and cross-lingual is today also a commonly used form.} is to build search engines that use a query expressed in one language (e.g., English) to find content that is expressed in some other language (e.g., French).  CLIR and Machine Translation (MT, automatically rendering the content of a document in another language) are closely linked; what we call MT is the use of translation technology to render documents readable, whereas CLIR is the use of translation technology to render documents searchable. So CLIR and MT are both built on the same foundation: {\em translation technology\/}. While our focus in this chapter is on retrieval of documents in one language using queries in another language, in practical applications the collection to be searched can potentially contain documents in many different languages, or even documents that themselves contain mixed languages, and the query too may contain terms from different languages.\footnote{Sometimes such cases have been referred to as Multilingual Information Retrieval (MLIR)~\cite{peters2012multilingual}, although that terminology is inconsistently used.  For example, many authors have used MLIR to refer to information retrieval systems such as Google that are able to handle many languages, even when no support for cross-language search is provided.}

\subsection{Some Use Cases}
The use cases for CLIR systems can be divided into two groups: (1) cases in which the user lacks proficiency in the language of the documents, and (2) cases in which the user can understand the documents, but prefers to use a different language for their query. These use cases, together with other factors such as relatedness of the languages, frequency of their interaction, collection size and format, or the expected number of queries, then influence the design choices which need to be considered when building a CLIR system. Here we describe several scenarios in which CLIR systems can be useful.

\begin{itemize}

\item In Web search, the natural emergence of one or more ``lingua franca'' languages such as English or Chinese results in an asymmetry in which much of the information is in some small number of languages, but many users prefer to (or must) interact with the Web in some other language.  In such cases, CLIR can provide access to information that would otherwise be unavailable.

\item Recall-oriented CLIR applications, which include searching in medical documents~\cite{Jimmy:2018,kelly:2016,kelly:2019,palotti:2017,Rosemblat2003,saleh:2020}, patents \cite{Lupu2017} and security applications~\cite{lincoln:2016}. The end users of such systems are often professionals for whom finding all documents on a defined topic is a high priority. It can be expected that the end users in such settings might have the ability to pay for high-quality human translation of the relevant documents. Even in such cases, however, automatic translations might provide a fast overview of the search results and they might also be helpful for deciding if individual documents require human translation. In such cases, a summary of a document in the language of the query might be useful.

\item CLIR applications can be expected to be particularly important in regions where multiple languages are frequently used, such as in countries that have more than one official language. This might include English/French in Canada, Dutch/French/German in Belgium, Spanish/Catalan/Basque in Spain, the 24 official languages in the European Union, or the 23 official languages in India.  It can be expected that substantial amounts of translated text would often be available in such settings, either through commercial practice or to meet legal requirements. Users of CLIR systems in such settings might be able to understand multiple languages, but they may be more fluent in, or more comfortable using, some of those languages. 

\item Question answering (QA) is a special case of Information Retrieval (IR) in which rather than simply identifying relevant documents to the user, the system seeks to use information from relevant documents to provide a direct answer to a question. Cross-Language Question Answering (CLQA) has received considerable attention among researchers~\cite{martino:2017,DBLP:conf/clef/GonzaloO04,loginova:2018,ruckle:2019} and with the recent attention to conversational search and chatbots it might be expected that this topic would attract continued attention in the coming years.

\end{itemize}

\subsection{The Three Waves of Information Retrieval}
We live today in the third wave of research on automated approaches to Information Retrieval (IR).  In the first wave, documents were typically described by controlled vocabulary descriptors (which is one example of what we today call metadata) and queries were expressed using the same descriptors.  These descriptors were typically organized hierarchically (e.g., {\em aircraft\/} is a narrower term than {\em vehicle\/}) in a thesaurus, and cross-language search could thus be supported by adding {\em entry vocabulary\/} in other languages (e.g., by linking the French free text term {\em avion\/} to the controlled vocabulary term {\em aircraft\/})~\cite{oard1998cross}.  The driving technology of that first wave was disk storage, and the driving need was to index large collections of published materials in libraries, and large collections of unpublished technical documents; the American Documentation Institute was an early venue for research on thesaurus-based access to large collections.\footnote{Today's successor to the American Documentation Institute is the Association for Information Science and Technology (ASIS\&T).}

In second wave IR (1990-today), documents are typically represented using the terms they contain (i.e., full-text indexing), searchers are no longer limited to thesaurus terms in their queries and instead can use any terms they wish, and retrieved documents are ranked based on the degree to which they are estimated by the system to satisfy the query. A driving force behind the emergence of the second wave was the need to provide search engines for the World-Wide Web, and the ACM Special Interest Group on Information Retrieval (SIGIR) emerged as a focal community for that work. 

In second-wave IR research, the usual approach was to represent queries and documents similarly, and then to compare the queries to the documents.  Examples of this included vector space and query likelihood models.  Recently, however, a third wave has emerged, with different representations for queries and for documents.  The defining technology for this new approach is the neural transformer architecture, through which query and document terms learn to attend to each other.  

Each of these waves of IR research has shaped a generation of CLIR research.  Our goal in this chapter is to present a concise view of CLIR as seen from the third decade of the twenty-first century.  It is not possible to review every detail, but neither is that necessary, since several useful overviews already exist.  Particularly notable in that regard is Nie's excellent survey, penned a decade ago~\cite{nie2010cross}.  Here, therefore, we pay particular attention to influential work published since Nie's 2010 survey. Along the way, our presentation supplements four other surveys that have been published since Nie's survey~\cite{dwivedi:2016,Kehinde:2018,zhang:2020,zhou:2012}.

\section{The Core Technology of CLIR}

CLIR research came of age in the second wave, when the key question was how queries and documents that were expressed using different languages could be represented similarly.  Because that work still shapes our view of CLIR, we first introduce how that was done.  We organize our story around three major questions: (1) `what to translate?', (2) `which translations are possible?', and (3) `how should those translations be used?'  

\subsection{What to Translate?}
\label{sec:what}
For CLIR, the question of what to translate exists at two levels.  First we can ask which items should be translated---the queries or the documents?  With that question answered, we must then ask how those documents or queries can be broken up into units (which we call terms) that can be translated.  First things first, we start with the question of whether it is the query or the documents that should be translated.

\subsubsection{The Query or The Documents?}\label{qd}
\hfill\\

Perhaps it seems obvious that it is the query that should be translated, since queries are short and documents are long.  Indeed, query translation is widely used in CLIR experiments, specifically because it is efficient.  But efficiency considerations may come out differently when the query workload is very high, as is the case in Web search.\footnote{The indexed Web contains 5.6 billion pages, but Google alone receives 5.6 billion queries per day.  At that rate, it doesn't take many days for the number of query words to become larger than the number of indexed document words.}  On the other hand, query translation has clear advantages when the goal is to cull through streaming content for documents that satisfy a standing query.
The point here is that the conventional wisdom that query translation is more efficient may or may not be true---much depends on the details of the task.

However, document translation has another potential advantage beyond the (relatively rare) cases in which it is the more efficient choice. 
The words in a document occur in sequence, and we have good computational models for leveraging such sequences to make more accurate translations. Of course, some queries include word sequences as well, but word sequences in queries are often very short, and thus possibly less informative. 
Modern translation models are trained using paired sentences from paired documents, and pairings of documents with their translations are much more widely available than examples of pairings of queries with their translations.
If people wrote queries like they wrote documents, this would not be a problem.  But they don't.  As a result, even when query translation would be more efficient, document translation might be more accurate.  

One further factor to consider when choosing between query translation and document translation is which direction yields the simplest index structure.  Query translation has advantages for applications in which there are many possible query languages, but only one document language.  In such cases, space efficiency argues for indexing in the document language, and thus translating the queries.  Symmetrically, document translation has advantages when all the queries are all in one language, but there are many document languages to be searched.  In such cases, space efficiency argues for indexing in the one query language.

Translation direction might moreover influence the translation quality. Translating into morphologically rich languages usually performs worse than translating into languages with simpler morphology, such as English. Translating from a language with no given word boundaries, such as Chinese or Japanese, might be harder than translating into such language. Translation direction thus might also have an effect on CLIR. Another obvious advantage to document translation is that if the translations produced at indexing time are designed to be read by people (not all translations used in CLIR are) then readers who are unable to read retrieved documents in their original language can be shown cached translations. So the picture is complex, with several factors to be considered when choosing between query translation and document translation.  But those are not the only two choices---there are two others.

The third choice is to translate both the queries {\em and\/} the documents. For example, if we want to use Navajo queries to search Burmese content we might translate both the Navajo and Burmese into English, simply because there are more language resources for Navajo-English and for Burmese-English than there are for Navajo-Burmese. 
If we broaden our concept of translation to include any mapping of an item (e.g., a word or a character sequence) from a representation specific to one language to some representation that is not specific to that language, then we can think of bilingual embedding (e.g.,~\cite{litschko2018unsupervised}) as a form of translation. 
With bilingual or multilingual embeddings, addressed in detail in Section~\ref{sec:embeddings}, we map from language-specific representations (in this case, sparse vectors that indicate the presence of specific terms in a specific language) to distributed representations (i.e., dense vectors) that can be thought of as language-independent representations of meaning. 
Here we have in some sense translated both the query and the documents; we have just translated them into an abstract representation rather than a representation associated with any single language.

As long as we're being exhaustive here, we should note that doing no translation at all is the fourth possibility.  When character sets are the same, sometimes proper names or loan words will match.  Although this is almost never a good idea on its own (with the possible exception of very close dialect differences such as British and American English), the presence of such felicitous ``translations'' can serve as useful anchor points around which more sophisticated and capable approaches can be constructed.  For example, McNamee and Mayfield used blind relevance feedback to enrich queries with related terms, thus producing a surprisingly good baseline for some language pairs to which other techniques could be compared~\cite{DBLP:conf/sigir/McNameeM02}.

Finally, we should note that when suggesting translating the query or translating the documents as possibilities, one need not limit themselves to just translating once.  Translation activity follows cultural contact patterns, and translation systems are most easily built for the language pairs in which there already is substantial translation activity.  For language pairs for which there is little cultural contact (e.g., Navajo and Burmese, as in our earlier example), it can be useful to translate through a ``pivot'' language (in this case, perhaps English) that has cultural contact with both. For example, to use Navajo queries to search Burmese content with query translation, we might first translate the query from Navajo to English and then from English to Burmese.  As \citet{DBLP:conf/cikm/BallesterosS03} have shown, applying query expansion each time translation is performed can serve to limit the adverse effect of cascading errors.

\subsubsection{Which Terms?}
\hfill\\

When we think of translation, we often think about translating words.  But the concept of a word is actually quite slippery.  Is \textit{poet} a word?  If so, is \textit{poets} a different word, or is it something else---a plural {\em form\/} of a word?  If we think of words as abstract objects that have forms, then \textit{poet's} and \textit{poets'} and even \textit{poet} are yet more forms of the same word, and \textit{poet} is its {\em root form\/}. 
But information retrieval people have a much simpler view of the world.  We don't call what we use in our searches words, but rather {\em terms},\footnote{The one exception is the so-called ``bag of words'' representation, which is of course in actuality a bag (in the set theory context) of terms.}
and a term is simply whatever we choose to index in an IR system.  This might be a token (in which case we need to say what choices our tokenizer makes!), it might be a stem (the part of the word which is shared across the inflected variants of this word)~\cite{resnik-etal-2001-improved}, or it might simply be a character sequence (which we call character n-grams).

Many other such choices are also made when choosing terms.  For example, terms can be in their original case or lowercased, diacritic marks can be retained or removed; substitutable characters can be retained as-is or normalized, and stopwords can be removed or retained.  The choices made on each of these points often differ from those used when human-readable translation is the goal, because information retrieval tasks (and thus information retrieval evaluation measures) often place greater emphasis on recall (i.e., on minimizing false negatives) than do machine translation tasks in which human readability is the goal.

None of this is specific to CLIR; these same issues arise in any IR application.  What is specific to CLIR, however, is that these choices affect not just the general retrieval approach, but also the translation functions that are indeed specific to CLIR.  In the simplest approaches to CLIR, we simply translate terms in one language into terms in the other language.  We can pay attention to word order (which often differs between languages) if our retrieval system leverages proximity or word order (or both), but often it may suffice to simply get the right terms in the other language.  So, for example, we could even translate from character n-grams in one language to character n-grams in the other~\cite{mcnamee2008textual}.

It is worth noting that we have used tokens, stems and character n-grams as examples, but there's more to making terms than just those three options.  For example, to our way of thinking in information retrieval, the statistical phrases used in Statistical Machine Translation (SMT) are also terms. So in CLIR, terms are not just what you index; they are also what you translate.  Note, however, that this introduces a tension, since longer terms limit translation ambiguity, whereas shorter terms offer more scope for improving recall.

\subsection{Which Translations are Possible?} \label{sec:which}

Once we choose which terms we need to be able to translate, our next questions are which translations are possible for each term, and (extending that further) how likely is it that each possible translation reflects the intended meaning?  There are at least four kinds of places we might look for the answer to that question:
\begin{enumerate}
    \item We might look translations up in some manually built lexicon.  In CLIR, this is referred to as dictionary-based translation (regardless of whether the lexicon was actually created from a human-readable dictionary).  Most often, this technique is applied to queries, and thus referred to as Dictionary-Based Query Translation (DBQT).  In practice, there are many kinds of manually built lexicons that might be mined; among the most common are bilingual phrase lists, bilingual dictionaries (which contain not just term mappings but also definitions, parts of speech, and examples of usage), multilingual thesauri (thus reusing technology from the first wave), or multilingual ontologies (e.g., titles of pages in different languages on the same topic mined from Wikipedia).
    Using actual bilingual dictionaries would make it possible to use part of speech tags as constraints, but accurate part of speech tags can be hard to generate automatically for short queries so this constraint has proven to be of less use than might be expected.  Some lexicons order alternative translations in decreasing order of general use, and that ordering can be exploited to avoid overweighting uncommon translations. 
    \item We might estimate translation probabilities from observed language use in one of three ways:  
    \begin{itemize}
    \item By far the most successful way of doing this in CLIR has been to mine parallel text (i.e., translation-equivalent passages; usually sentences) to learn translation mappings that include a probability for each translation. This can be done explicitly using Statistical Machine Translation (SMT) models~\cite{xu-weischedel-2000-cross} or Neural Machine Translation (NMT) models~\cite{ZbibZKHDHJRZSM19}, or it can be done implicitly using a complete SMT or NMT system that produces one-best or n-best results. Of course, the relative prevalence of translations in the parallel text that is used to learn the probabilities should be similar to that expected in the collection to be searched; learning from translations of tractor manuals might not be wise when the goal is to search poetry. 
    \item An alternative source would be comparable corpora (i.e., collections of documents on similar topics that are not translations of each other), which can be used to learn plausible translations.  When links between pairs of documents on the same language are available, these mappings can either be learned explicitly~\cite{sheridan1996experiments} or implicitly~\cite{littman1998automatic,ruder2019survey}. Indeed, it is even possible to learn implicit mappings with no links between document pairs at all by leveraging regularities in usage that are preserved across languages~\cite{chen-cardie-2018-unsupervised}.  These comparable corpus techniques are limited by corpus size, however, since orders of magnitude larger corpora are needed for comparable than for parallel corpora to get similarly good results for relatively rare words. This issue of modeling the translations of rare words well can be particularly important for CLIR because in many settings relatively rare words are relatively common in queries. 

    \item Finally, it is also possible to mine collections of mixed-language documents for translations.  One approach to this is to focus on inline definitions that explain the meaning of a term in one language using terms in another language~\cite{li2006translation}.  Such an approach to writing is used to introduce new technical terms in some settings.
    \end{itemize}
    \item We might also generate (or implicitly recognize) ``translations'' using an orthographic or phonetic transcription algorithm.  Buckley, writing somewhat tongue in cheek, famously observed that for purposes of information retrieval, French can be thought of as misspelled English and `corrected' using a spelling correction algorithm~\cite{buckley2000using}.  More generally, transliteration can be performed based on spelling (essentially, character-level translation) or pronunciation, it can be generative (forward transliteration), inferential (backward transliteration), or implicit (cognate matching)~\cite{chen-etal-2018-report}.  Transliteration can be particularly useful for proper names, but it is also a potentially useful fallback any time translations are not available from lexicons or from corpus-based methods.
    \item Finally, the original source of all translation knowledge is human intelligence, so when all else fails we might turn to the user for help.  For example, an interactive system might leverage limited evidence of possible translations to engage the searcher in a dialog regarding how best to translate a query term~\cite{ogden1999keizai}, or systems might show searchers snippets that include untranslatable terms in the hope that the searcher would be able to interpret the term in context.
\end{enumerate}


\subsection{How to Use those Translations?} \label{sec:psq}

Much of the early work on dictionary-based query translation involved simply substituting one document-language term for each query-language term.  Since many terms are homonymous,\footnote{People often say polysemous when they mean to refer to different meanings for terms that are written identically, but when speaking precisely polysemy refers to closely related meanings---often not a problem in IR---whereas homonymy refers to unrelated meanings.} by which is meant that they have more than one quite different meaning, that naturally leads to the question of which translation(s) should be used.  A common approach, both in early CLIR work and in MT, has been to look to other terms in the query in an effort to disambiguate the intended meaning of each query term.  However, one term might have both distantly and closely related translations (e.g. the Spanish word {\em banco\/} can be translated into English as bank, bench, or pew), 
so a better question (at least when more than one of the translations is plausible) would be which translation\underline{s} should be used.  That naturally leads to the question of how several possible translations for a single term should be used, and that turns out to be the right question in general (since a single known translation is simply a special case of that more general question).

Initial attempts at using multiple translations, such as stuffing all of the translations into one very long query, proved to be problematic.  Indeed, even careful attention to the balance of weights between query terms did not work particularly well.  The fundamental problem was that ranking functions tend to give more weight to query terms that are rare, so it was the rarest translations that were dominating the results---exactly the opposite of what you would want.  Still today, it is not uncommon to see unbalanced query translation inappropriately used as a baseline in CLIR experiments.  A better approach, based on closer coupling of translation and retrieval, was ultimately introduced by Pirkola~\cite{DBLP:conf/sigir/Pirkola98}.  In Pirkola's Structured Query (SQ) method, the key idea was to think of translation as synonymy:
every time any synonym (here, any translation) of a query term was seen in a document, the count for the query term in that document was incremented.  
With this insight in hand, it was a simple matter for Pirkola to use the synonym operator in the Inquery information retrieval system, a generalization of Inquery's query-time stemming, to perform CLIR.  

Meanwhile, a parallel line of work on CLIR had developed in which people were simply cascading Machine Translation (MT) and IR to create CLIR systems.  This too usually just substituted one word for another, but now with the MT system making the choice of what that single replacement should be.  CLIR researchers soon realized that more could be gained by using the internal representations of the MT system, however, to identify synonyms.  Thus dictionary-based and MT-based CLIR researchers were on convergent paths.

The unification of CLIR and MT that occurred in the late 1990's led to a further extension to the structured query method, asking not just which translations were possible, but how likely each document term was to have been translated into a query term.  Note the direction of the conditional probability here; we are interested in the probability that a query term is a translation of a document term, not the other way around.  We can then ask, for every document, what would be the expected counts of the query terms if the document had been written in the query language.  If $e_{i,j}$ is the expected count of query term $e_i$ in document $j$ and $f_{k,j}$ is the actual count of document term $f_k$ in document $j$ then we get:
\begin{equation}
e_{i,j} = \sum_{k \in j} p(e_{i}|f_{k}) f_{k,j},
\end{equation}
where $p(e_{i}|f_{j})$ is the probability that document term $f_k$ (e.g., banco) would translate to query term $e_i$ (e.g., pew).  This probability might be estimated in many ways, but by far the most popular approach when parallel (i.e., translation-equivalent) text is available has been to perform term-level alignment and then compute the maximum likelihood estimate for the probabilities (i.e., the approach known in machine translation as IBM model 1)~\cite{DBLP:conf/acl/McCarley99}.  

This approach to using estimated translation probabilities was first tried with document ranking functions based on unigram language models, but the formulation is general and can be applied with any ranking function.
In general, ranking functions in information retrieval are based on three elements of a term-by-document matrix: (1) an element statistic based on the number of occurrences of a term in a document (for representing the``aboutness'' of a document), (2) a row statistic based on the number of documents in which a term occurs (for representing the specificity of a term), and (3) a column statistic based on the number of terms in each document (that can used to compute a density from the element statistic).  For brevity, these are typically called term frequency (TF), inverse document frequency (IDF), and length, respectively. It is useful to remember, however, that in practice what is often meant are not raw counts but rather monotonic functions of those counts.

With that in mind, it is easy to see how the same approach can be used to estimate document frequency.  Defining $k$ as the number of distinct terms in the document language, $|f_{k}|$ as the number of documents containing term $f_k$, and $|e_{i}|$ as the expected number of documents in which query term $e_i$ would have appeared (if the documents had been written in the query language), we get:
\begin{equation}
|e_{i}| = \sum_{k} p(e_{i}|f_{k}) |f_{k}|
\end{equation}

This formula actually overestimates $|e_{i}|$ somewhat (because two terms that translate to $e_{i}$ might be found in the same document), but the error has been shown experimentally to typically be inconsequential~\cite{DBLP:conf/sigir/DarwishO03}.  Alternatively, if a suitably large and representative side collection is available in the query language, $|e_{i}|$ can be estimated directly from that side collection. For length, it typically suffices to directly use the length as computed in the document language.  It is well known that different languages can use more or fewer words to express the same content, but for documents of any reasonable length the scaling factor is fairly close to a constant, and most of the widely used ranking functions produce rankings that are insensitive to constant factors.  So adjustments to the document length are not typically required.  Darwish and Oard called this approach of separately estimating query-language term frequency vectors and a query-language document frequency vector from the observed document-language statistics Probabilistic Structured Queries (PSQ), since it extends Pirkola's SQ approach by allowing partial mappings~\cite{DBLP:conf/sigir/DarwishO03}. The approach can also be seen as an extension of the approach that language modeling researchers had earlier introduced for the case in which only term frequency statistics needed to be mapped across languages~\cite{xu-weischedel-2000-cross}.

PSQ is essentially vector translation; you start with either the TF vector or the DF vector in the document language, multiply that vector by a translation probability matrix, and get a TF or DF vector in the query language.  At that point, ranking proceeds as if the documents had been written in the query language in the first place.  The key point is that TF and DF vectors should be translated separately; precombining those two factors (e.g., using TFIDF or BM25 term weights) before translation does not work nearly as well.\footnote{The reason for this is that IDF is perhaps better thought of as a measure of the selectivity of a query term than as a measure of selectivity of any particular document-language translation of that query term.}  The only twist is that these ``translated'' TF or DF vectors are no longer vectors of integer counts, but rather are now vectors of real-valued (i.e., partial) term counts.  This is typically associated with some spreading of the counts across more terms, which has implications for both effectiveness and efficiency.  If implemented at query time (i.e., indexing only the document language TF and IDF statistics), then the index will be no larger than would be the case for monolingual IR, but query time will scale linearly with the average translation fanout. If implemented at indexing time (i.e., indexing query-language term weights) then the index size will increase linearly with the average translation fanout, but the only upward pressure on query time would be the increased time to load the larger postings files from secondary storage (e.g., hard disk or solid-state drive)~\cite{oard2002translation}. Empirical results indicate that adverse effects on retrieval effectiveness from that spreading are often balanced by gains in (average) effectiveness from the expansion effect of the translation mapping (which can usefully add synonyms).  Adverse effects on efficiency can be limited by pruning the set of possible translations, although overly aggressive pruning can adversely affect effectiveness.

This line of development left one key question unresolved: how translation should be done when the ranking function makes use of term proximity, term order, or both.  The problem was that the initial approach to PSQ translated only decontextualized terms.  The first to tackle that problem were Federico and Bertoldi, leveraging the observation that translation can change word order, but that long-distance reordering is less common~\cite{DBLP:conf/sigir/FedericoB02}.  They therefore experimented with a variant of what has subsequently come to be called the Sequential Dependence (SD) model~\cite{metzler2005markov} in which both original and reversed word orders were treated as equally informative.

Work on PSQ still continues, most recently by~\citet{DBLP:journals/ir/RahimiMS20}, who show that further improvements can be obtained in some cases by balancing discrimination values (e.g., IDF) calculated in the query and the document languages.

\section{Updating Nie: CLIR since 2010}

If a CLIR counterpart to Rip Van Winkle were to have fallen asleep in 2010, only to awake in 2022, the thing that surely would most impress them is the tremendous energy around neural methods.  Neural networks offer ways of learning nonlinear models that have the potential to improve over simpler linear and hand-engineered nonlinear models~\cite{haireport}.  Widespread adoption of Rectified Linear Units (ReLU) over the last decade has made it possible to train much deeper models, and this in turn has led to a blossoming of new model architectures.  Applications of these techniques to CLIR have faced three challenges: (1) Initial work on Neural MT focused principally on one-best translation for human readability rather than on tuning for retrieval tasks, (2) neural retrieval models are data intensive, and present training data has largely focused on monolingual applications, and (3) neural methods are computationally intensive, thus placing a renewed premium on efficiency.  In this section we adopt key aspects of the structure of the Chapter 2 to review advances in CLIR since 2010. 

\subsection{What to Translate: Parts of Words} \label{sec:parts}
As described above, systems for handling text as tokens must choose how to segment that text into terms. This task has received particular attention among Machine Translation (MT) and Automatic Speech Recognition (ASR) researchers, who have developed new representations that can also be useful for retrieval tasks. Byte-Pair Encoding (BPE)~\cite{sennrich-etal-2016-neural} and Wordpiece~\cite{6289079} are now widely used in MT. Both these methods use subword units to deal with rare words, thus mitigating the out of vocabulary problem. In BPE, subword splitting is first trained on a collection so that a subword vocabulary of a pre-specified size is created. Wordpiece works similarly, but the decision about which subwords to generate is made based on the maximizing language model log likelihood rather than word frequency as in BPE. Wordpiece is especially widely used in BERT models (see below). In contrast to these models, which operate on individual tokens, Sentencepiece~\cite{kudo-richardson-2018-sentencepiece} operates on the sentence level by first joining all the words in the sentence, and then re-segmenting that aggregated character sequence by inserting a special token at the segmentation points.

\subsection{Which Translations: Bilingual and Multilingual Embeddings}
\label{sec:embeddings}

The third wave of CLIR (characterized by different representations for queries and for documents) is built from three key innovations: (1) some way of creating dense vector representations in which terms with similar meaning are represented by similar vectors, (2) deep neural architectures that can be trained using back-propagation, and (3) the learned self-attention of the neural transformer architecture.  The first of these emerged initially in second-wave IR research.  Initial work on creating dense vector representations for terms dates to the introduction of Latent Semantic Indexing (LSI) in 1988~\cite{DBLP:conf/sigir/FurnasDDLHSL88}. At the time, linear algebra, specifically a truncated Singular Value Decomposition (SVD), was used to compute a dense representation for each term (known as a singular vector).  Today, neural autoencoders use non-linear models to do the same thing: compute a dense representation for each term.  These dense representations capture an embedding of the original high-dimensional term space in a lower-dimensional vector space.  Given a fixed vocabulary (e.g., from BPE), they can be trained in advance, thus reducing the term-level embedding process to a simple table lookup.

It did not take long for researchers to realize that bilingual embeddings could be created that assigned similar dense representations to terms with similar meanings, regardless of their language.  Initially these bilingual embeddings were also constructed using linear algebra~\cite{littman1998automatic,littman1998learning}; today such representations can also be learned using autoencoders~\cite{bonab:2020,vulic2015monolingual}.  
Three broad classes of techniques have been proposed for creating bilingual term embeddings: (1) pseudo-bilingual, (2) projection-based, and (3) unsupervised.  For a detailed treatment of these approaches generally, we refer readers to a survey on this topic by \citet{sogaard2019cross}.  Here we focus on those parts of the story that involve CLIR.

\citet{landauer1991statistical} were the first to propose the \textit{pseudo-bilingual} approach.  The key idea in this approach is to create a set of mixed-language documents from a comparable corpus using explicit links between the documents (as discussed above in Section \ref{sec:which}), and then to learn embeddings from that collection.  Specifically, they concatenated English and French versions of the same document, and then applied a truncated SVD to learn a dense representation for terms in both languages.  Because the truncation tended to preserve meaning while suppressing the effect of term choice, the approach learned similar vectors for terms in different languages that had similar meaning.  Twenty-five years later, \citet{vulic2015monolingual} did something similar using a neural autoencoder and document-aligned comparable corpora from Wikipedia and Europarl. Instead of an SVD, they trained a skip-gram \cite{DBLP:journals/corr/abs-1301-3781} model on documents created by concatenating an English document and a comparable document in Dutch, and then shuffled the terms. When used with an unusually wide window, this allows the skip-gram model to take into account bilingual context to generate an embedding for each term. However, randomly shuffling the terms might lead to sub-optimal choice of bilingual context for certain terms. Improving over random shuffling, \citet{bonab:2020} used a bilingual dictionary to guide the process of interleaving terms from the two languages so that terms that are putative translations of each other would be adjacent in merged sentences. This makes it likely that a valid translation will appear in the neighborhood context of a term, which will be then used to learn better representations for both that term and its translation. In their work, they used sentence-aligned parallel text, and merged on a sentence-by-sentence basis.

\citet{littman1998learning} was the first to propose the \textit{projection-based} approach for CLIR, where independently learned monolingual embeddings are aligned using a learned linear transformation such that known translations (e.g., from a biligual dictionary, as described in Section \ref{sec:which}) yield similar representations. This approach is now very widely used, typically being applied to align monolingual embeddings that were created using autoencoders.  Because fairly large dictionaries are now available for many language pairs, the projection-based approach is easily scaled to create multilingual embeddings that today can represent terms from more than 50 languages~\cite{ammar2016massively,joulin-etal-2018-loss,DBLP:conf/iclr/LampleCRDJ18}. In recent work with embeddings learned using neural autoencoders, \citet{bhattacharya2016using} used linear regression to learn a transformation between a Hindi and English monolingual embeddings using Hindi-English translation dictionary. Similarly, \citet{litschko2018unsupervised} mapped source and target embeddding spaces into a single shared space by learning two projection matrices, one for each language, to create a new joint space that is different from that used for either language~\cite{DBLP:conf/iclr/SmithTHH17}. In that work, they compared three ways of creating projection matrices: Canonical Correlation Analysis (CCA) \cite{faruqui2014improving}, minimizing the Euclidean distance between the projected vector and the target vector \cite{DBLP:journals/corr/MikolovLS13}, and maximizing a retrieval-oriented measure they call Cross-Domain Similarity Local Scaling (CSLS) \cite{joulin-etal-2018-loss,DBLP:conf/iclr/LampleCRDJ18}. In multilingual CLIR, \citet{bhattacharya-etal-2016-query} used two different approaches to learn multilingual embeddings, a projection-based method in which Bengali and Hindi were mapped to an English embedding space, and the pseudo-bilingual approach proposed by \citet{vulic2015monolingual}, extended to multiple languages. 
 
Recently, there has also been some work on \textit{unsupervised} approaches to learning bilingual embeddings using projection-based techniques \cite{artetxe-etal-2018-robust,hoshen-wolf-2018-non,DBLP:conf/iclr/LampleCRDJ18}. While supervised approaches rely on sources such as bilingual dictionaries to learn alignments, unsupervised approaches instead follow a two-step process: 1) build an initial seed dictionary from monolingual vector spaces using some unsupervised approach, and 2) iteratively refine that dictionary to add more pairs that will eventually be used to learn better embeddings. Although there exist several unsupervised approaches to learn the initial dictionary, the key idea in the second step is to project one embedding space onto another and then to use some of the near neighbors in the aligned spaces as the added dictionary entries. The induced dictionary can be iteratively refined using Procrustes analysis \cite{schonemann1966generalized}, and the refined dictionary can then be used to learn better bilingual embeddings. This does, however, assume that the vector spaces are isomorphic, which needs not be true for languages that are typologically different, and this approach has been shown not to work well in such cases \cite{vulic2019we}.  Litschko et al. \citep{litschko2018unsupervised,litschko2019evaluating} used these unsupervised approaches to perform fully unsupervised CLIR in language pairs involving limited training resources. However, the effectiveness of fully unsupervised approaches was found to be rather limited when compared to supervised methods by \cite{vulic2019we}. 

One problem with embeddings learned using either linear algebra or autoencoders is that homonomy distorts the representation, since the model has no way of knowing if this {\em banco\/} is a financial institution or a wooden bench in a park.  The advent of neural methods using self-attention made it possible to produce context-sensitive embeddings, with the embedding for {\em banco\/} differing, depending on its surrounding terms~\cite{devlin2018bert,peters-etal-2018-deep}. The resulting Bidirectional Encoder Representations from Transformers (BERT) architecture \cite{devlin2018bert} has since been further extended (e.g. RoBERTa~\cite{liu2019roberta}, XLNet~\cite{yang2019xlnet}, ELECTRA~\cite{clark2020electra}, XLM~\cite{lample2019cross}), some of which support more than 100 languages. We should emphasize that this use of BERT and its derivatives to produce contextual embedings differs from the third-wave use of BERT that we describe in more detail below for direct query-document matching. 

All the techniques discussed so far for producing cross-lingual embeddings produce generic representations for queries and documents that need to be tuned for the CLIR task.
An alternative is to learn representations that are tailored for CLIR, as had already been done with for representations tailored for monolingual retrieval~\cite{huang2013learning,zamani2018neural}. The advantage of these ``representation-based'' learning approaches is that the representations of the documents can be precomputed, and during retrieval the score on which ranking is based can then be more rapidly computed (more on this in Section \ref{sec:neural-ir}). The key challenge then lies in learning specialized representations of queries and documents that are useful for this task. \citet{gupta2017continuous} learned CLIR-specific representations for queries and documents using a transfer learning approach based on a parallel corpus. First, they learned a representation model similar to DSSM~\cite{huang2013learning} by using a monolingual retrieval collection in English to learn representations such that similarity computed between representations of the queries and representations of the relevant documents are maximized, whereas similarity computed between representations fo the queries and representations of the non-relevant documents is minimized. Using a trained monolingual model, they extended it to the cross-language setting with an English-Hindi parallel corpus. To do so, the monolingual model was applied to the English side of the parallel text, and a cross-lingual model was trained on Hindi side. The goal was maximize the similarity of the English representation computed using the monolingual model and the Hindi representation computed using the cross-lingual model for a given translation pair. In a somewhat different approach, \citet{li2018learning} employed an adversarial learning framework using a Generative Adversarial Network (GAN)~\cite{goodfellow2014generative} to learn a CLIR-specific representation by incorporating both monolingual and cross-lingual matching signals. Their model takes as input three sources: raw queries, documents, and translated queries (in this case, translated using Google Translate). They incorporated several constraints in their model to learn task-specific representations: 1) a monolingual constraint that brings the representations of documents and translated queries closer, 2) a cross-language constraint that brings the representations of documents and raw queries closer, 3) a translation constraint that brings the representations of raw and translated queries closer, and 4) an adversarial constraint that produces source-agnostic representations. 

Instead of using an off-the-shelf multilingual contextual language model to generate generic embeddings, ~\citet{litschko2021cross,litschko2021evaluating} explored the use of multilingual contextual models fine-tuned on sentence-level parallel corpora (e.g., LASER~\cite{yang-etal-2020-multilingual}, m-USE~\cite{artetxe2019massively}, LaBSE~\cite{feng2020language}) to produce embeddings specialized for semantic similarity. Naturally, these models work better for matching queries with sentences than with longer documents. While the existing multilingual models are pretrained on generic tasks such as Masked Language Model (MLM), Translation Language Model (TLM) and Next Sentence Prediction (NSP), another line of work explored designing custom pretraining tasks that are tailored for CLIR.~\citet{yu2021cross} introduced Query Language Modeling (QLM), which is a variant of TLM with two differences. While TLM relied on sentence-level parallel-corpora, QLM uses the section titles of Wikipedia in one language as query and the relevant section in another language as the relevant document. Instead of masking tokens in both the source and target side of parallel sentences, as in TLM, QLM only masks tokens in the source side.~\citet{yu2021cross} further introduced a Relevance Ranking (RR) task that produces a higher ranking score using the contextualized embeddings of the concatenated sequence of query and relevant document than a sequence of query and non-relevant document. ~\citet{yang2022c3} introduced C3, a continual pretraining recipe which generalized Yu et al.'s RR task to include several non-relevant documents as hard negative samples. This framework, known as contrastive learning~\cite{chen2020simple}, has been shown to work well in many domains. C3 further introduced a token-level similarity task, which is part of the ColBERT~\cite{Khattab20a} framework, to promote cross-lingual alignment.

\subsection{Which Translations: MT}
MT has advanced quite rapidly over the past decade, building upon the key innovations in distributed word representations and deep learning technologies.  For high-resource language pairs, this has resulted in neural systems convincingly surpassing the performance of the statistical MT systems that had dominated MT research activity in the first decade of this century. This started with the advent of attention-based neural methods \cite{bahdanau2014neural} built on an underlying encoder-decoder architecture \cite{neco1997asynchronous}. The key idea behind the attention mechanism is to leverage the fact that models can pay attention to different parts of the source sentence while decoding each target token. Later on, the introduction of subword units (described above in Section \ref{sec:parts}) addressed the problem of translating rare words. Ultimately, these two factors, attention and subword units, proved to be the key ingredients.

As mentioned above in Section \ref{sec:what}, query translation is often a preferred approach owing to its computational efficiency in experimental settings. However, the key challenges lie in 1)~overcoming the ambiguity introduced by short word sequences in queries, and 2)~mismatch between the style of writing for the queries to be translated and the style of the parallel sentences used for training the MT models (as described in Section~\ref{qd}). Prior work has tried to tackle the style mismatch issue by creating an in-domain dataset of queries paired with their translations and using it to train SMT/NMT systems. In the context of CLIR, \citet{nikoulina:2012} first explored this idea for building a SMT system using a dataset consisting of human translated queries provided by CLEF. \citet{huquery} built a small manually-curated in-domain dataset consisting of popular search queries, augmenting it with large amounts of parallel out-of-domain sentences to train a NMT model.
As a different approach, \citet{yao2020exploiting} utilized click-through data from a multi-lingual e-commerce website\footnote{\url{https://www.aliexpress.com/}} to mine query translation pairs. 
Another way to minimize the style mismatch effect is to force the MT model to to incorporate terms from the retrieval collection in the translation of the query. This has been explored in SMT by using the target retrieval collection to train a language model, which helps in translating the query in a way that better matches the documents \cite{nikoulina2013domain}. 

Similarly, \citet{bi2020constraint} trained a neural model by constraining the translation candidates of the query to the set of document terms ranked highly by TFIDF. The set of documents used for this was selected using click-through data for each query. In a somewhat different approach, \citet{sarwar-etal-2019-multi} tried to solve the style mismatch problem using a multi-task neural model that optimized for improving the query translation jointly with a relevance-based auxiliary task.  The goal of the auxiliary task was to restrict the NMT decoder to produce translations using terms that are equally likely in both the retrieval collection and the MT training corpus. The auxiliary task involved learning a relevance-based embedding model on the content of the top-ranked document from the retrieval corpus, using the target language sentence from the parallel corpus as the query. Since these word embeddings are shared across the NMT and relevance models, that forces the NMT decoder to choose translations that occur in both the retrieval collection and the parallel corpus. 

Document translation, on the other hand, has the advantage of the additional context present in the longer documents. Attention-based encoder-decoder NMT models leverage the context present in both the source and the target side to produce context-dependent translations, in contrast to dictionary-based or PSQ techniques that rely on context-independent term translation. However, owing to the computational overhead involved in translating the entire document collection (as described above in Section \ref{sec:what}), there has been less work exploring neural models for document translation than there has been for query translation.
However, full encoder-decoder models are only one point in a larger design space.  Relaxing the contextual constraint to apply only to the document being translated, and not to the sequence of terms that are generated as possible translations, leads to an approach for generating multile translation hypotheses for each term called the Neural Network Lexical Translation Model (NNLTM)~\cite{ZbibZKHDHJRZSM19}. Applying constraints only on the source side implies that the translation(s) of the document terms are assumed to be independent of each other. While this would affect the fluency of translations intended for human readability, the principal goal of a translation model for CLIR is to generate alternative translations, and not necessarily to properly order all those translations. This is very similar to the case of PSQ relying on the word alignments generated from the SMT system rather than using the language model to help select and order those translations. For each document term, NNLTM uses the neighboring terms within a window to produce the contextualized translation alternatives. In order to do so, while training, NNLTM relies on the alignments from a parallel corpus generated using word aligners, and uses an aligned term in the document language along with the context words surrounding it term to predict the translation in the query language.

Ignoring other qualities of query or document translation described in section~\ref{sec:what}, such as speed, a natural question to ask is which approach is most effective? Contrary to our expectation, recent work from \citet{saleh:2020} found that translating queries actually works better than translating documents while retrieving texts from a medical domain. However, the queries were translated from European languages to English, while the documents were translated from English to other European languages, which is potentially a harder problem. \citet{DBLP:conf/acl/McCarley99} also tried to answer the question of whether the query or the documents should be translated, finding that neither approach was a consistent winner.

If the quality of MT and CLIR are interconnected, the subsequent question is how a machine translation system can be tuned to achieve better CLIR retrieval effectiveness. One commonly used approach to integrate retrieval signals with MT is to (re)-rank the translation alternatives produced by an MT system in order to maximize a retrieval objective. This was first attempted by \citet{nikoulina:2012}, who tuned their SMT system to produce translations that directly optimize Mean Average Precision (MAP). Since then, 
\citet{sokolov2014learning} have focused on optimizing SMT decoder weights to maximize a retrieval objective that leads to producing translations geared towards better retrieval. \citet{saleh:2016:b} have since built a feature-based reranking model that reranks different alternatives produced by an MT system and then selects the translation that maximizes a retrieval objective. 

\subsection{Using Translations: Neural IR} \label{sec:neural-ir}

Now that we have different ways of generating bilingual embeddings that are either generic or task-specific, as discussed in Section \ref{sec:embeddings}, the next question to ask is how to use them for CLIR. One possibility, which we have already mentioned, is to use \textit{embedding similarity} between query and document representations. One way to create such representations is to create an aggregated query vector from the embeddings of individual query terms and an aggregated document vector from the embeddings of the individual document terms. Some similarity measure such as the cosine can then be used to compute a retrieval score for each document, with documents sorted in decreasing score order. \citet{vulic2015monolingual} explored different forms of aggregation, including unweighted and weighted summation of word embeddings, using IDF as weights for the weighted summation. They found that IDF-weighted summation worked better than unweighted summation.

A second approach for incorporating embeddings into a retrieval system is to use similarity in a bilingual embedding space of terms from different languages as a basis for term translation. We refer to this approach as \textit{embedding-based translation}.  For example,~\citet{bhattacharya2016using} used the k most similar terms from the document language as translations of a query term. This is similar to query expansion, but in a cross-language setting. In a similar approach, \citet{litschko2018unsupervised} introduced a model that they called Term-by-Term Query Translation (Tbt-QT) which, as the name suggests, uses embedding-based translation of each term. Specifically, they used a single cross-language nearest neighbor for every query term. As might be expected from the experience with pre-neural CLIR methods, this turns out to be sub-optimal when compared with using more than one nearest neighbor~\cite{bonab:2020}. 

A third approach, which has now been well studied in monolingual IR, uses embeddings as input to custom neural architectures that are optimized for relevance and that are fast enough to perform full-collection neural ranking on modestly large collections. In one prominent line of work on monolingual information retrieval known as ``full-interaction-based'' methods, query and document terms are encoded using a single neural network initialized with Word2Vec monolingual representations \cite{DBLP:journals/corr/abs-1301-3781}, and the model learns interactions between those representations to maximize a relevance objective (DRMM~\cite{guo2016deep}, KNRM~\cite{xiong2017end}, or PACRR~\cite{hui2018co}). The work of \citet{10.1145/3397271.3401322} extends these full-interaction-based approaches to CLIR. These models are initialized with fastText bilingual embeddings aligned using a dictionary \cite{joulin-etal-2018-loss}, which is type of projection-based bilingual word embedding (as described in Section \ref{sec:embeddings}).

Subsequent research in monolingual retrieval has focused on switching from context-independent Word2Vec embeddings to the context-sensitive BERT-based representations described above in Section \ref{sec:embeddings} to initialize the neural ranking models. These third-wave context-sensitive Transformer-based architectures have achieved results superior to the best previously known techniques for ranking documents with respect to a query in monolingual applications~\cite{Khattab20a,lin:2019b,McAvaney20a}. ColBERT~\cite{Khattab20a} introduced a two-stage retrieval approach that relies on learned embeddings to produce an initial set of documents for reranking. These embeddings were trained MaxSim, a late-interaction-based mechanism that finds the document token embedding that is closest to a specific query token embedding. The key idea is that query and document embeddings can be computed separately; that allows us to precompute and store the document embeddings in a nearest neighbor index. During inference, the first-stage retrieval system computes a query embedding on-the-fly and then uses approximate nearest neighbor techniques to find the documents with terms that are closest to the token embeddings of the query. In the next stage, these documents are ranked using the one query token whose contextual embedding is most similar to that of each query term. While initially developed for monolingual applications, ColBERT has been generalized to the CLIR setting~\cite{li2021learning,nair2022transfer}. One variant, ColBERT-X~\cite{nair2022transfer}, used the translations of relevant passages in MS MARCO~\cite{bonifacio2021mmarco} to build a CLIR model to retrieve documents in non-English languages using English queries. ~\citet{li2021learning} showed that a generalization of ColBERT to CLIR also works when retrieving English documents using non-English queries.

Full-interaction-based ranking models that make connections between each query term and every document term can be computationally expensive when ranking all of the documents in a very large collection such as the Web. This leads to the fourth way of employing embeddings in CLIR, \textit{cascade-based approaches}. These involve running an efficient recall-oriented ranker first to get an initial ranked list, which is then re-ranked using a more computationally expensive model \cite{dai2019deeper,nogueira2019passage,yilmaz2019cross}. However, the retrieval effectiveness of cascade re-ranking approaches is limited by by the quality of the initial ranked list produced by a fast term-matching recall-oriented ranker.  \citet{zhang-2019} extended the cascade re-ranking approach to the cross-language setting and found similar improvements as in monolingual retrieval, especially on low-resource languages. \citet{jiang-etal-2020-cross} fine-tuned a pretrained multilingual BERT model on cross-language query-sentence pairs constructed from parallel corpora in a weakly supervised fashion and applied that model to perform re-ranking for CLIR. \citet{shi2019crosslingual} used a transfer learning approach, applying a retrieval model trained on a large collection in English to retrieve documents from collections in other languages. \citet{litschko:2022} aimed on reducing the amount of training data needed to fine-tune the model on various languages by using  Adapters and Sparse Fine-Tuning Masks.

\subsection{Fusion}
\label{sec:fusion}

Different approaches to IR have different strengths and weaknesses.  This inherent diversity suggests that there is sometimes potential for improving search results by combining different approaches. CLIR systems have even greater potential for benefiting from diversity than do monolingual IR systems because of the additional potential for diversity that translation resources, and ways of using those translation resources, introduces.  Combining multiple sources of evidence has a long heritage, and the inclusive name for the general approach is fusion~\cite{Croft:2002,culpepper:2018,wu:2012}. In IR we distinguish between two classes of fusion, \textit{early} fusion (in which evidence from multiple sources is combined by one or more components of the full system before results are generated) or \textit{late} fusion (in which the results of separate IR systems are combined).  
Late fusion is often referred to as system combination.  In general, late fusion can be employed with systems that search different collections, but here we focus only on cases in which all systems search the same collection.  

The earliest work on system combination for CLIR dates to 1997 when, as noted above, \citet{DBLP:conf/acl/McCarley99} found that neither query translation nor document translation was a clear winner. He went on to test a late fusion combination between the two approaches, finding that the combination of the two approaches yielded the best results.  McCarley had endeavored to hold every other aspect of the system design constant, but later experiments by \citet{braschler:2004} and by \citet{kazuaki:2006} demonstrated benefits from combining more diverse systems as well.

Early fusion of translation resources has also proven to be successful in CLIR.  Perhaps the best known case is the fusion of parallel sentences with translation pairs from a bilingual lexicon when training machine translation systems.  The architecture for this is simple; machine translation systems are trained on pairs of translation-equivalent sentences, and term pairs from a bilingual lexicon are simply treated as very short sentences.  Because the term distribution in bilingual lexicons are not as sharply skewed in favor of common terms as is naturally occurring parallel text, this approach can help to overcome a natural weakness of parallel text collections (limited size) and by helping to more reliably learn translations of relatively rare terms.

One challenge that arises in both early and late fusion is that the evidence to be merged may be represented differently.  This occurs, for example, when merging translation probabilities (e.g., from statistical alignment) with translation preference order (e.g., from a bilingual dictionary) in early fusion, or when merging document scores from different retrieval systems in late fusion.  In such cases, some form of score normalization is needed. This problem has been extensively studied in monolingual fusion applications, and approaches like CombMNZ~\cite{lee:97} with sum-to-one normalization that work well in monolingual settings also seem to work well in CLIR~\cite{nair:2020b}.

\citet{ture-2012} conducted early fusion experiments with three ways of estimating term translation probabilities (one with language model context, one with term context, and one with no context), finding substantial improvements over using the best of those three representations alone. \citet{ture2014learning} later extended this approach to include learned query-specific combination weights for a similar set of three translation probabilities (two with language model context and one with no context), finding further improvements. \citet{han:2019} explored an alternative approach, learning late fusion combination weights without supervised (retrieval effectiveness) training by instead using data programming to estimate system weights.  \citet{nair:2020b} compared early and late fusion using two approaches (one with language model context, one without), finding similar improvements over the best single system from both approaches. Finally, \citet{DBLP:conf/aaai/LiLWBLL0Y20} experimented with using cross-attention for early fusion between CLIR (for product descriptions) and product attribute matching in an e-commerce application.

One particularly important consideration when using neural methods is that neural methods excel at generalization for cases in which there is adequate training data, but they have more difficulty modeling rare phenomena.  Because queries often tend to skew toward more specific, and thus less common, terms, a commonly used approach has been to combine neural and non-neural results.  Good results have been reported for that approach in CLIR as well, both with early fusion~\cite{bonab:2020,huang2021mixed} and with late fusion~\cite{schamoni-2014}.

\subsection{Cross-Language Speech Retrieval}
\label{multimodal}

We have so far described retrieval of content (and queries) represented as digital text. But CLIR can be also applied to content that is spoken in a language different from that of the queries. When doing CLIR on recorded speech, a straightforward approach would be to first perform Automatic Speech Recognition (ASR) and then to treat the resulting text as a text CLIR problem. This is the approach that was used by all of the participants in the first four shared-task evaluations on cross-language speech retrieval~\cite{marcello:2003,marcello:2004,oard2006overview,pecina2007,white:2006}.  This can work well when ASR accuracy is high, but there are still many settings (e.g. conversational speech, or speech recorded under difficult acoustic conditions) in which one-best ASR transcripts exhibit substantial word error rates.  As the throughput of computing resources has dramatically increased over the last two decades, it has thus become more common to take advantage of the more complex internal representations of ASR systems.

Internally, ASR systems typically represent what might have been spoken as some form of confusion network (for what might have been spoken between fixed time points) or lattices (for what might have been spoken between arbitrary time points). These representations might be phonetic or word-based.  For word-based confusion networks that are aligned to the one-best word boundary, equations (1) and (2) in Section 2.3 are easily extended to incorporate an additional multiplicative factor for transcription probabilities estimated from ASR confidence scores.  As \citet{nair:2020} found, this can yield better CLIR results than would be achieved from one-best ASR. 
\citet{yarmohammadi-2019} have also reported a beneficial effect from concatenating multiple recognition hypotheses from utterance-scale (i.e., sentence-like) confusion networks.   

Phonetic approaches to cross-language speech retrieval have also been explored.  Because translation is typically performed on words, not phonemes, this approach is better matched to a query translation CLIR architecture.  This is the approach that was tried in the very first reported work on cross-language speech retrieval~\cite{sheridan1997}, in which query terms were first translated using a comparable corpus, then rendered phonetically, and matched with phoneme trigrams created for the spoken content using an ASR system.

Finally, it is worth noting that there has been a spurt of recent activity on cross-language speech retrieval as part of a large multi-site research program in the United States known as MATERIAL~\cite{Zavorin:2020}. Two things distinguish MATERIAL from much of the prior work.  First, the program experimented with nine less commonly studied languages (Bulgarian, Farsi, Georgian, Kazakh, Lithuanian, Pashto, Somali, Swahili and Tagalog).  Using results from that program, \citet{Rubino:2020} found that differences in retrieval effectiveness were much better correlated with the amount of available training data for ASR and MT than with the phonetic or syntactic similarity of the query and content languages.  Second, the goal of a MATERIAL system was not to rank documents, but rather to select a set of relevant document~\cite{bbn:2020:b,aqwv}.  Set selection is a challenging task, but it is a task that is well matched to the needs of speech retrieval applications.  In text retrieval, the searcher might easily skim a ranked list of results, and in CLIR they might use MT to help them with that task.  In speech retrieval, however, the linear nature of audio means that skimming can take longer, and the speech translation systems that would be needed to support that task for cross-language speech retrieval are less accurate than are current text translation systems.  In some applications, it may therefore be useful to think of cross-language speech retrieval as a triage application, selecting documents that will then require time from expert speakers of the language.  Systems developed for MATERIAL (e.g.,~\cite{boschee-etal-2019-saral,bbn:2020,Galuscakova:2020}) are thus useful as exemplars for complex cross-language text and speech retrieval systems that can draw on the full range of techniques described above.

\subsection{Building Real Systems}

CLIR research tells us what can be built, and it offers some prescriptions for how best to build it. 
Compared with the architecture of the IR system, CLIR pipeline contains extra steps which help us crossing the language barrier. They might be done at different spots of the pipeline, depending on the approach. For example embedding-based approaches require using multilingual model instead of the monolingual one and translation-based approaches require running a translation before retrieval. This additional complexity might be also reflected in indexes, which might need to store additional information such as the probability of the translation in the PSQ method.
Many details remain to be worked out when CLIR techniques are employed in real systems.  Here we highlight two of those issues: language-specific processing, and accommodating multilinguality.

 In research systems, we often strive language-independent methods.  In an earlier era, a chapter on CLIR might have devoted considerable attention to the characteristics of specific languages. In recent years, however, CLIR has been evaluated on text and speech collections in a substantial number of language pairs without yet finding pairs for which the same basic approaches would not work. Methods such as Byte-Pair Encoding and Wordpiece, for example, diminished the need for language specific tokenization. Practical systems, by contrast, can still benefit substantially from language-specific processing. As the language of the queries and documents differs, CLIR systems naturally involve two (or more) sets of language-specific issues. For example, in English, stemming is normally applied only to suffixes.  In Arabic, by contrast, stemming must pay attention to both prefixes and suffixes, and the agglutination means that Arabic stemming will benefit even more from recursion than would English stemming. Dialect variations in some languages can introduce a whole new level of complexity, and it is not uncommon to find multiple dialects in the same collection.  It is not uncommon to find unicode characters used incorrectly, particularly when the visual appearance of different characters is similar.  And even when correctly encoded, differences in transliteration conventions (or, more properly, the lack of a single transliteration convention) can result in a substantial variety of representations for the same word when that word has been transliterated from a different character set.  Transliteration of queries can also introduce considerable variation, as keyboards on different devices may limit the characters that can easily be generated.  The number of language specific issues is so great that one is reminded of the old adage that success is 10\% inspiration and 90\% perspiration.

Although our focus throughout this chapter has been on cases in which queries are expressed in one language and documents are expressed in one other language, the world is not that simple.  Many collections include mixed languages, and many documents are multilingual for the simple reason that they include so-called loan words from other languages. Multilinguality poses a number of challenges. Detecting the predominant language in a document is a fairly easy task, but there are cases in which the language of a query might only be reliably guessed if some information about the person asking the query is known. For example, \textit{Gift} means marries in Swedish, a present in English, and poison in German. 
The challenges are not limited to the queries or the documents, the system itself has choices to make.  For example, if potentially relevant documents in many languages can be found, the system might opt to process and index them in the same way or language specifically. It could also display several ranked lists and allow the user to choose, or it could try to merge the results and display one ranked list. 
  
\section{Evaluation}

As with IR generally, shared task evaluations and test collections have been driving forces in the development of CLIR, helping researchers to identify and collaborate on emerging problems, providing the training and evaluation data needed to make technical progress, and fostering comparability across research sites. 

\subsection{Shared-Task Evaluation}

Nie provides an overview of the shared-task evaluations organized through 2010. At that time, only the TREC book had been published~\cite{voorhees2005trec}; since then, edited volumes that review two of the other major CLIR evaluation venues (CLEF and NTCIR) have appeared~\cite{peters2012multilingual,sakaievaluating}. The Text REtrieval Conference (TREC) was the venue for the first cross-language retrieval tasks. Early work on CLIR was conducted as a side activity associated with work on non-English collections (in Spanish and Chinese).  The main focus of the TREC CLIR tasks between 1997~\cite{vorhees:2000} and 1999~\cite{voorhees:1999} was on CLIR in multiple languages. It was there that a key decision that has influenced CLIR test collection design was first sorted out: the topics (from which the queries are generated) should be translated as natural expressions in the target language, so as to more faithfully represent what a real searcher might pose to a CLIR system.  Early in the new century TREC returned to bilingual tasks involving Chinese or Arabic~\cite{gey:2000,gey:2001,oard:2002} and then again in 2022 as in the NeuCLIR task focused on Chinese, Persian, and Russian. The focus on CLIR for European languages moved to the newly formed Cross-Language Evaluation Forum (CLEF) in Europe. As its name suggests, a range of CLIR-related tasks were organized at CLEF, including ad-hoc CLIR and domain specific CLIR, interactive CLIR, Cross-Language Spoken Document Retrieval and Cross-Language Retrieval in Image Collections. 

Starting in 2012, CLEF embraced broader range of text and multimedia processing tasks and was renamed as the Conference and Labs of the Evaluation Forum. CLIR-related tasks at CLEF after 2012 include eHealth, which has included multilingual search in medical data, and Multilingual Cultural Mining and Retrieval (MC2). The CLIR part of the MC2 Lab task \cite{mc2} was focused on microblog search. Short critiques of movies in French were used as the queries. The main focus was on finding related microblogs in order to provide a diverse summary of a movie based on comments and movie information in French, English, Spanish, Portuguese and Arabic. 
CLIR has been part of the eHealth tasks since 2014 \cite{ehealth:2014}. The task was in 2014 focused on search by the general public rather than medical professionals. The collection contained one million of medical documents. Queries were originally created for the monolingual task in English and were subsequently translated into  Czech, French and German. Though the task in the following year \cite{ehealth:2015} was still oriented towards the general public, queries were created to imitate users searching in the database using observed medical symptoms. Arabic, Farsi and Portuguese were added to the languages of the queries. The ClueWeb 12 B13 collection was used as the collection in the following years and the queries were mined from Web forums to better imitate realistic conditions \cite{ehealth:2016,ehealth:2017}. The list of languages also changed, as Hungarian, Polish and Swedish were supported in addition to  German, French and Czech.
In 2018~\cite{ehealth:2018}, webpages from the Common Crawl corpus were used as the collection and queries that had been issued by  nonprofessional users of the public HON\footnote{\url{https://www.hon.ch/en}} search service were used. 

CLIR tasks involving Japanese and some other East Asian languages were first organized at what is now called the NII Testbeds and Community for Information access Research (NTCIR) evaluations between 2002 and 2007. NTCIR tasks related specifically to CLIR in that period included Cross-Language Q\&A, Advanced Cross-Lingual Information Retrieval, Question Answering, Patent Retrieval, and Cross-lingual Link Discovery. Since 2010, CLIR was part of two tasks: GeoTime (between 2010 and 2011), and Crosslink (between 2011 and 2013). In the GeoTime task \cite{gey:2010,gey:2011}, the search was constrained by geographic and temporal constraints using `when' and `where' queries. English and Japanese queries were used to search in English and Japanese news. The main focus of the Crosslink task \cite{crosslink:2011,crosslink:2013}, was automatic linking of related content in different languages.

The Forum for Information Retrieval (FIRE) was created in 2008 to focus on South Asian languages.  CLIR tasks at FIRE have included Ad-hoc Cross-language Document Retrieval (between 2010 and 2013), SMS-Based FAQ Retrieval (between 2011 and 2012), Cross-Language Indian News Story Search (CL!NSS) (in 2013), and Mixed Script Information Retrieval (between 2015 and 2016). Similarly to CLEF, the focus of FIRE has gron over time to include a broader range of multilingual access problems such as language identification, event and information extraction, offensive content and hate speech detection and fake news detection. 

Recently, CLIR was also the focus of the OpenCLIR challenge,\footnote{\url{https://www.nist.gov/itl/iad/mig/openclir-challenge}} which used a Swahili test collection from the MATERIAL program.

\subsection{Test Collections}

Shared task evaluations provide test collections and evaluation frameworks to their participants, often with an eye toward making those test collections available to future researchers. We focus in this section on test collections which are presently publicly available; an overview is in Table~\ref{tab:data}. We list both CLIR test collections and test collections used in closely related problems, such as cross-language question answering. In addition to shared task evaluation venues, which were traditionally the leaders in providing test collections, companies such as Microsoft and Facebook have recently provided access to a range of test collections, especially those oriented towards question answering. Those collections are typically much larger than the collections used in shared task evaluations, which is particularly important for training data-hungry neural techniques. In contrast to this, the collections such as the Large-Scale CLIR Dataset\footnote{\url{http://www.cs.jhu.edu/~kevinduh/a/wikiclir2018}} by \citet{sasaki:2018}, CLIRMatrix\footnote{\url{http://www.cs.jhu.edu/~shuosun/clirmatrix}} by \citet{sun-duh-2020-clirmatrix}, and the XQA\footnote{\url{https://github.com/thunlp/XQA}} open-domain question answering collection by~\citet{liu-etal-2019-xqa} 
were created from Wikipedia. Though these collections can also be useful for training neural models, their synthetic nature makes them more suitable for pretraining than for task-specific fine tuning. We therefore omit such collections from our summary. We also omit the early TREC collections, which were well covered in Nie's review. Among question answering test collections, we focus only on those with broad language coverage; more test collections focused on single languages also exist (e.g. MMQA~\cite{mmqa}).

\begin{savenotes}
\begin{sidewaystable}
\tiny
\centering
\setlength{\tabcolsep}{0.4em}
\begin{tabular}{|p{0.24\linewidth}|p{0.06\linewidth}|p{0.15\linewidth}|p{0.15\linewidth}|p{0.20\linewidth}|p{0.13\linewidth}|}
\cline{1-6}
\bf{Collection} & \bf{Published} & \bf{\#Queries} & \bf{\#Documents} &\bf{Languages} & \bf{Availability}\\
\hline
The CLEF Test Suite for the CLEF 2000-2003 Campaigns – Evaluation Package~\cite{clef:2000,clef:2001,clef:2002}\footnote{Apart from the data, the CLEF Initiative topics, experiments, pools and performance measures from 2000 to 2013 are also available at \url{http://direct.dei.unipd.it/}. }  & 2001-2003 & 40/50/50/60 (depends on the year and language) & 44-454k (depends on the language) & English, French, German, Italian, Spanish, Dutch, Swedish, Finnish, Russian, Portuguese & ELRA\footnote{\url{https://catalogue.elra.info/en-us/repository/browse/ELRA-E0008/}}\\
\hline
TDT 1-5~\cite{tdt1,tdt2,tdt3} & 2001-2006 & 25/96/80/250 topics & 16-98k in total (depends on the year and collection) & Arabic, Chinese, English & LDC\footnote{\url{https://catalog.ldc.upenn.edu/LDC98T25}}\footnote{\url{https://catalog.ldc.upenn.edu/LDC2001T57}}\footnote{\url{https://catalog.ldc.upenn.edu/LDC2001T58}}\footnote{\url{https://catalog.ldc.upenn.edu/LDC2005T16}}\footnote{\url{https://catalog.ldc.upenn.edu/LDC2006T19}} \\
\hline
NTCIR-3-6 CLIR: IR/CLIR Test Collection~\cite{ntcir3,ntcir4,ntcir5,ntcir6} & 2002-2007 & 80/60/50/50 topics (depends on the year) & 10-900k (depends on the year and language) & Chinese, English, Japanese, Korean & Online application\footnote{\url{https://www.nii.ac.jp/dsc/idr/en/ntcir/ntcir.html}}\\
\hline
CLEF Question Answering Test Suites (2003-2008) – Evaluation Package~\cite{clefqa2003,clefqa2004,clefqa2005,clefqa2006,clefqa2007,clefqa2008} & 2003-2008 & 200 topics per year & 55k-454k (depends on the language) & Bulgarian, Dutch, English, Finnish, French, German, Italian, Portuguese, Romanian, Spanish & ELRA\footnote{\url{http://catalog.elra.info/en-us/repository/browse/ELRA-E0038/}}\\
\hline
NTCIR CLQA (Cross Language Q\&A data Test Collection)~\cite{ntcirqa:2005,ntcirqa:2007} & 2004-2007 & 200 topics per year & 10-900k (depends on the year and language) & Chinese, English, Japanese & Online application\footnote{\url{https://www.nii.ac.jp/dsc/idr/en/ntcir/ntcir.html}}\\
\hline
CLEF AdHoc-News Test Suites (2004-2008) – Evaluation Package~\cite{adhoc2004,adhoc2005,adhoc2006,adhoc:2007,adhoc2008} & 2004-2008 & 50 topics per year & 17-454k (depends on the language) & 14 languages & ELRA\footnote{\url{https://catalogue.elra.info/en-us/repository/browse/ELRA-E0036/}}\\
\hline
CLEF Domain Specific Test Suites (2004-2008) – Evaluation Package~\cite{domain:2004,domain:2005,domain:2006,domain:2007,domain:2008} & 2004-2008 & 25 topics per year & 20-303k (depends on the language) & English, German, Russian & ELRA\footnote{\url{https://catalogue.elra.info/en-us/repository/browse/ELRA-E0037/}}\\
\hline
GALE Phase 1-4 \cite{strassel-etal-2006-integrated, babko-malaya-2008-annotation,babko-malaya-etal-2010-evaluation} & 2005-2009 & depends on the collection & depends on the collection & Arabic, Chinese, English & LDC\footnote{\url{https://www.ldc.upenn.edu/collaborations/past-projects/gale/data/gale-pubs}}\\
\hline
NTCIR ACLIA (Advanced Cross-Lingual Information Retrieval and Question Answering)~\cite{aclia2008,aclia2010} & 2008-2010 & 100 topics per language each year & 249k-1.6M (depends on the year and language) & Chinese, English, Japanese & Online application\footnote{\url{https://www.nii.ac.jp/dsc/idr/en/ntcir/ntcir.html}}\\
\hline
FIRE Information-Retrieval Text Research Collection~\cite{fire:2008,fire:2010,fire:2012} & 2008-2012 &  50 topics per language each year & 95-500k (depends on the year and language) & Assamese, Bengali, English, Gujarati, Hindi, Odia, Marathi, Punjabi, Tamil, Telugu & Online application\footnote{\url{http://fire.irsi.res.in/fire/static/data}}\\
\hline
NTCIR Crosslink (Cross-lingual Link Discovery)~\cite{crosslink:2011,crosslink:2013} & 2010-2013 & 28 / 25 topics & 202k-3.6M (depends on the year and language) & Chinese, English, Japanese, Korean & Online application\footnote{\url{https://www.nii.ac.jp/dsc/idr/en/ntcir/ntcir.html}}\\
\hline
CLEF eHealth 2014-2018~\cite{ehealth:2014,ehealth:2015,ehealth:2016,ehealth:2017,ehealth:2018} & 2014-2018 & 50/50/67/300/50 topics (depends on the year) & 23k / 1M / 5.5 M (depends on the year) & Arabic, Czech, English, Farsi, French, German, Hungarian, Polish, Swedish & Differs by collection\footnote{\url{https://sites.google.com/site/clefehealth/datasets}}\\
\hline
Czech Malach Cross-lingual Speech Retrieval Test Collection & 2017 & 118 topics &  353 interviews (592 hours) & Czech, English, French, German, Spanish & CC-BY-NC-ND 4.0\footnote{\url{https://lindat.mff.cuni.cz/repository/xmlui/handle/11234/1-1912}}\\
\hline
Extended CLEF eHealth 2013-2015 IR Test Collection \cite{ehealth} & 2019 & 50/50/166 (depends on the year) queries & 1.1M & Czech, English, French, German, Hungarian, Polish, Spanish, Swedish & CC BY-NC 4.0\footnote{\url{https://lindat.mff.cuni.cz/repository/xmlui/handle/11234/1-2925}}\\
\hline
MLQA~\cite{lewis2019mlqa} & 2020 & around 5k Q/A pairs per language & 2-6k articles per language & English, Arabic, German, Spanish, Hindi, Vietnamese, Chinese & CC-BY-SA 3.0~\footnote{\url{https://github.com/facebookresearch/mlqa}}\\
\hline
MKQA: Multilingual Knowledge Questions \& Answers~\cite{mkqa:2020} & 2020 & 10k Q/A pair per language & - & 26 languages & CC BY-SA\footnote{\url{https://github.com/apple/ml-mkqa}}\\
\hline
XQuAD~\cite{artetxe-etal-2020-cross} & 2020 & 1190 Q/A pairs & 240 paragraphs & 11 languages & CC-BY-SA 4.0~\footnote{\url{https://github.com/deepmind/xquad}}\\
\hline
XTREME~\cite{hu2020xtreme} & 2020 & differs by task & differs by task & 40 languages & Apache License v2.0\footnote{\url{https://github.com/google-research/xtreme}}\\
\hline
XOR QA~\cite{asai2020xor} & 2020 & 2-3 Q/A pairs per language & - & Arabic, Bengali, Finnish, Japanese, Korean, Russian, Telugu & CC BY-SA 4.0\footnote{\url{https://nlp.cs.washington.edu/xorqa}}\\
\hline
HC4: HLTCOE CLIR Common-Crawl Collection\cite{hc4} & 2022 & 60 topics & 0.5M - 5M (depends on the language) & Chinese, English, Persian, Russian & Online\footnote{\url{https://github.com/hltcoe/HC4}}\\
\hline
\end{tabular}%
\caption{Overview of CLIR collections. If the collection contains more than 10 languages, the number of languages is shown.}
\vspace{-3em}
\label{tab:data}
\end{sidewaystable}
\end{savenotes}

One important caveat when using these collections is that older test collections may not include judgments for relevant documents that are found by newer systems~\cite{lin2020pretrained}.  This could produce the paradoxical result that better systems might get worse scores.  Moreover, CLIR systems are subject to advances not just in the core IR technology, but also in the translation technology that they use, and (for cross-language speech retrieval) in the ASR technology that they use.  A number of evaluation measures that are robust to random omissions of relevance judgments are known (e.g., infAP and bpref), but the problem with new methods tested on older collections is that the omissions may well not be random.  Users of older test collections would thus be wise to check the number of highly ranked unjudged documents in their result sets, and in the result sets for the baselines to which they compare, and to caveat their results if substantial differences in the prevalence of unjudged documents is observed.

Finally, we note that the full scope of IR evaluation design issues in IR arise in CLIR as well.  Most notably, comparison to weak baselines, as noted generally by \citet{armstrong:2009}, seems particularly problematic in CLIR.  Hopefully, this chapter helps to address that problem going forward.  Moreover, test collections on which CLIR results are reported are often smaller than those for IR more generally, which also means that greater attention to issues of experiment design and reporting~\cite{Fuhr:2017,Sakai:2020} continue to be particularly important.  Finally, the development of standard CLIR systems that are widely available lags well behind the situation for IR more generally~\cite{diaz:indri,anserini}, thus posing additional challenges for reproducability.

\section{Future Directions}

As the development of CLIR techniques is closely tied with developments in IR and natural language processing, it can be expected that trends in these areas will be reflected in CLIR. Especially as neural approaches have only recently risen to prominence in IR, further growth of those techniques in CLIR can also be expected. With dominance of neural machine translation techniques in recent years, it can be expected that closer integration of machine translation and information retrieval techniques, such as joint training and tuning, will be further explored. Apart from deep neural networks, other machine learning concepts, such as reinforcement learning, imitation learning, bandit algorithms or metalearning may also find their way into CLIR. 

The challenge of interaction design for IR also exists in CLIR. There has been far less reported research on interaction design for CLIR than on monolingual retrieval, and the work that has been done (e.g., \cite{chenjun:2018,DBLP:conf/hicss/OgdenD00,DBLP:journals/umuai/PetrelliN06,DBLP:conf/jcdl/ZhangPKOS07})
points to a number of open research questions. Interactive CLIR systems could, for example, use information about preferred document languages to decide which documents to rank highly when relevant documents exist in several languages. Another challenge is monetization of interactive search. If the search engine chooses to present advertisements to the user, they must be in a language understandable by the user. Moreover, the advertisement should be connected with the topic of the search and search results, and they need to be for products or services that can actually be provided to that searcher.

One natural direction for CLIR research is continuation of its integration with multimedia retrieval. With the rise of voice assistants (e.g., 39.4\% of Internet users in US used a voice assistant at least once a month in 2019)\footnote{\url{https://mobidev.biz/blog/future-ai-machine-learning-trends-to-impact-business}} and thus the increasing popularity of conversational search, cross-language conversational search may become another emerging topic. As language support by voice assistants is still somewhat limited,\footnote{\url{https://www.globalme.net/blog/language-support-voice-assistants-compared}} one may expect that CLIR in conversational search might become particularly important for languages with fewer speakers. In such cases, the relevant content might only exist in a language different from the language of the speaker, and the voice assistant will need to decide whether and how to present answers from such content to the speaker. Cross-language conversational search will thus face new challenges, such as determining the probability that the top relevant document really contains the answer to the user's question, determining whether the answer differs in different languages, or exploring  how to present the answer to the user if it only exists in a language different from the query language. 

One notable gap in the CLIR literature is the application of CLIR to what has been called microtext: short informal text such as that found on Twitter, Short Message Service (SMS) text messages, or text chat.  Such sources pose unusual challenges such as colloquial abbreviations (e.g., `u r fine'), intentional misspellings (e.g., `pleeeeeeze`), and device-influenced transliteration conventions.  Mixed-language context is common, and many types of information occupy the same channel (e.g., Twitter contains professional announcements, thoughts or activity descriptions, and a good deal of spam).  While there has been some experimentation with CLIR from informal content~\cite{bagdouri2014clir}, and some work on translation of Twitter content that could perhaps be applied to CLIR (including one Twitter MT paper that actually uses standard CLIR techniques for data augmentation~\cite{jehl2012twitter}), the dearth of work on CLIR from microtext seems somewhat surprising.  It thus seems reasonable to anticipate that this topic may attract attention in the coming years.

Language-based search in speech and video is still an under-explored area, given the vast quantities of content that might be found in such sources. Monolingual search in podcasts recently received attention thanks to increasing commmercial interest in that content,\footnote{\url{https://podcastsdataset.byspotify.com}}
and podcasts offer a natural source of content for cross-language speech retrieval. CLIR for spoken content is today seen as a straightforward extension to CLIR, but of course speech is richer in some ways than text (e.g., with evidence for mood, emotions, or speaker accent). It seems plausible that at least some of this potential will one day be explored in a CLIR context.

Image content might be easier to present to speakers of different languages then speech or text, as noted by~\citet{wrro4524}. Indeed, caption-based cross-language image search has been the focus of tasks at both Image CLEF and in the iCLEF interactive task. 
Work on this problem has continued in recent years with CLIR for metadata that describes both images \cite{menard:2015,li:2019}  and video \cite{Khwileh2016}.  The recent emergence of systems that learn to generate captions directly from image content opens new directions for image retrieval that has yet to be explored in CLIR applications.

Another emerging topic in  Information Retrieval is fairness and bias. Similarly to neural networks and conversational search, these were identified as emerging IR topics by the 2018 SWIRL report~\cite{allan2018swirl}, and they might reasonably be expected to ultimately be as important in CLIR research as in monolingual IR. \
A CLIR system might, for example, be biased towards returning documents from a certain category (e.g., research papers from universities might be returned more often than the research papers from government organizations). Especially in multilingual CLIR, one might ask if the system is (or should be!) biased towards returning documents from particular languages. So, as with many things, CLIR simply adds an additional dimension of complexity.

As we have sought to illustrate in this chapter, CLIR is yesterday's problem---much of the groundwork has already been laid.  It is also today's problem, since the Web provides the potential to access a vast cornucopia of knowledge that, without CLIR, most of its users simply would not be able to find.  And it is tomorrow's problem, since wherever IR research goes, CLIR seems sure to follow!

\section*{Acknowledgements}

This work has been supported in part by NSF grants 1618695 and 1717997, and by the Office of the Director of National Intelligence (ODNI), Intelligence Advanced Research Projects Activity (IARPA), via contract FA8650-17-C-9117. The views and conclusions contained herein are those of the authors and should not be interpreted as necessarily representing the official policies, either expressed or implied, of ODNI, IARPA, or the U.S. Government. The U.S. Government is authorized to reproduce and distribute reprints for governmental purposes notwithstanding any copyright annotation therein.

\bibliography{bibliography}

\end{document}